\documentclass[onecolumn,preprint,superscriptaddress,english,prl,titlepage,longbibliography]{revtex4-2}

\usepackage[colorlinks=true,urlcolor=blue,citecolor=blue,linkcolor=blue]{hyperref}
\usepackage{makecell}
\usepackage{enumitem} 
\usepackage{graphicx}
\usepackage{dcolumn}
\usepackage{bm}
\usepackage{amsmath,amssymb,amsthm,mathrsfs,amsfonts,dsfont}
\usepackage{balance}
\usepackage{color}
\usepackage{array}
\usepackage{tabularx}
\usepackage{longtable}
\usepackage{color}
\usepackage{float}
\usepackage{indentfirst}
\usepackage{txfonts}
\usepackage{booktabs} 
\usepackage{multirow}

\usepackage{placeins}
\usepackage{verbatim}

\usepackage{physics}
\usepackage{algorithm} 
\usepackage{algorithmic}
\usepackage{subfloat}

\begin{document}

\title{Correcting a noisy quantum computer using a quantum computer}
\author{Pan Zhang}
\email{panzhang@itp.ac.cn}
\affiliation{Institute of Theoretical Physics, Chinese Academy of Sciences, Beijing 100190, China}
\begin{abstract}
Quantum computers require error correction to achieve universal quantum computing. However, current decoding of quantum error-correcting codes relies on classical computation, which is slower than quantum operations in superconducting qubits. This discrepancy makes the practical implementation of real-time quantum error correction challenging.
In this work, we propose a decoding scheme that leverages the operations of the quantum circuit itself. Given a noisy quantum circuit $ A $, we train a decoding quantum circuit $ B $ using syndrome measurements to identify the logical operators needed to correct errors in circuit $ A $. The trained quantum circuit $ B $ can be deployed on quantum devices, such as superconducting qubits, to perform real-time decoding and error correction.
Our approach is applicable to general quantum codes with multiple logical qubits and operates efficiently under various noise conditions, and the decoding speed matches the speed of the quantum circuits being corrected. We have conducted numerical experiments using surface codes up to distance 7 under circuit-level noise, demonstrating performance on par with the classical minimum-weight perfect matching algorithm.
Interestingly, our method reveals that the traditionally classical task of decoding error-correcting codes can be accomplished without classical devices or measurements. This insight paves the way for the development of self-correcting quantum computers.
\end{abstract}
\maketitle


An ideal quantum computer holds the potential to achieve exponential speedups over classical computation in tasks such as prime factoring~\cite{factor1997,FTneed_2HowFactor20482021}. However, quantum computers are susceptible to noise at the physical level~\cite{preskillnisq2018quantum}, which must be corrected to ensure accurate computation results~\cite{qec1fowlerintroduction2012,qec2latticesurgery2012,qec3SingleshotError2019,qec4XZZXSurface2021,qec5qccBB2024,qec6_DynamicallyGeneratedLogical2021,qec7_ConstantOverheadFaultTolerantQuantum2023,qec8_landahl83FaulttolerantQuantum2011,qec9_fowler2DColorCode2011,qec10_bombinTopologicalSubsystemCodes2010,ducloscianci2010renormalization,bravyi2014Efficient,Bombin_2012,Bombin2007,gauge}. Quantum error-correcting codes tackle this challenge by integrating multiple physical qubits into a smaller number of logical qubits, encoding the logical information through redundant measurements on ancilla qubits~\cite{exp2_menendez26ImplementingFaulttolerant2023,reichardt2409demonstration,exp4_hyperproduct2024,exp5_2024,exp6_2024,exp7_color_20211,zhaoRealizationErrorCorrectingSurface2022,bluvstein2023Logical,googlequantumai2023Suppressing,acharya_quantum_2025,googlerep}. When an error occurs, nontrivial measurement results are triggered and used to infer how the logical information has changed, allowing for appropriate corrections. This inference process, known as decoding, is considered a classical statistical inference problem. A wide range of decoding algorithms~\cite{dennis2002Topological,  higgott2021pymatching, higgott2023Sparse, wu2023fusion,correlatedmatching2013,believematching2023,Roffe2020, Panteleev2021degeneratequantum,cao2025exact} have been proposed to address this problem, including minimum-weight perfect matching (MWPM), union-finding, belief propagation with ordered statistics for decoding (BPOSD)~\cite{412683, 924874, Roffe2020}, tensor network methods~\cite{googlequantumai2023Suppressing,chubb2021General}, and neural-network decoders~\cite{bausch2024learning,gnd,qecgpt}. Despite the availability of numerous decoders and accelerations using modern computational power like GPUs and FPGAs, accurately correcting practical quantum codes under actual circuit-level noise at a speed aligned with the operation of quantum devices with superconducting qubits remains challenging, making real-time decoding difficult.
In this work, we propose to address the decoding problem using a quantum circuit rather than classical computational devices, thereby aligning the decoding speed with the operation speed of the quantum circuit to be corrected. In the following sections, we will first formulate the decoding problem within the context of quantum error correction codes. We will then introduce our method and demonstrate its performance through numerical experiments using surface codes up to distance 7 under circuit-level noise. Finally, we will extend our proposed method to explore the possibility of self-correcting quantum circuits, which can correct errors without requiring ancilla qubit measurements or classical electronic interactions.
\begin{figure*}[htb]
    \centering
    \includegraphics[width=0.9\linewidth]{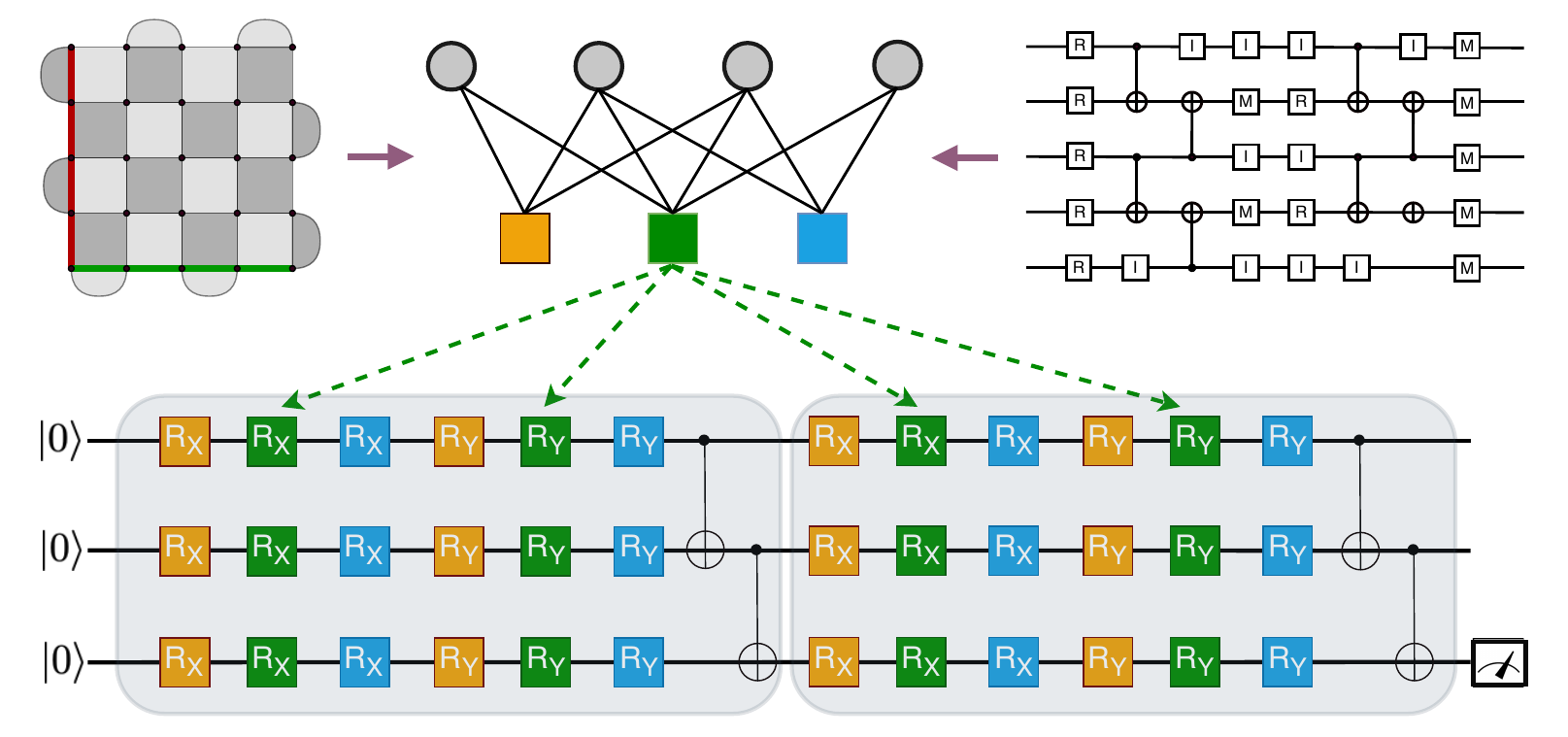}
    \caption{\textbf{Top}: Pictorial illustration (middle) of the Tanner graph generated from various codes. The top left panel depicts a rotated surface code under qubit noise, while the top right panel shows a distance-3 repetition code under circuit-level noise. In the notation, R stands for reset, M for measurement, and I for idling. Their corresponding Tanner graphs are illustrated in the top middle panel, with only three syndrome measurements depicted in different colors. \textbf{Bottom}: A decoding circuit composed of three qubits. The decoding circuit consists of two blocks of unitary operations. Each block contains three X-rotation gates and three Y-rotation gates, along with two Controlled-Z gates. The color of each rotation gate corresponds to the syndrome measurement shown in the top middle panel.  The parameters of the rotation gates are $\theta_i x_i$ where $\theta_i$ is a learnable parameter and $x_i=\{0,1\}$ denotes the syndrome.}
    \label{fig:circuit}
\end{figure*}

\paragraph{Quantum codes and decoding}
We introduce the Tanner graph representation of a quantum code, as shown in the top middle panel of Fig.~\ref{fig:circuit}. The circles in the graph represent error sources, each of which has a certain probability of experiencing an X or Z error, depending on the error model. For qubit-error models, the error sources are the qubits themselves; an example of the surface code is depicted in the top left panel. For circuit-level noise models, the error sources could be specific regions within the circuit; an example of the repetition code under circuit-level noise is illustrated in the top right panel. The colored squares in the Tanner graph represent XOR measurements on the error sources, which are constructed using measurements on ancilla qubits. The results of these measurements, known as the syndrome, indicate whether an odd number of errors has occurred in a particular set of error sources. If no error occurs, the syndrome would be trivial (all zeros). However, if an error does occur, certain measurements will be triggered and report a non-trivial syndrome. For $ m $ measurements, there are $ 2^m $ possible syndromes $\gamma = \{\gamma_1, \gamma_2, \cdots, \gamma_m\} \in \{0,1\}^m$. 

The task of decoding is to infer the logical operation $\beta$ made by the error, given the syndrome $\gamma$. The minimum-weight decoders convert the problem to finding the most likely error consistent with the syndrome. Here, weight is related to the negative log-probability of the error being generated in the error model. For the surface code with independent X and Z noises, minimum-weight decoding can be achieved very efficiently, and the resulting algorithm, such as the minimum-weight perfect matching algorithm (MWPM), has been widely applied. 

Minimum-weight decoders often compromise decoding accuracy because they overlook the degeneracy of errors. Many errors belong to the same logical sector and are equivalent due to their identical commutation relations with logical operators. To elaborate, the number of logical sectors is $4^k$, where $k$ is the number of logical qubits. A logical sector can be identified using a vector $\beta = \{\beta_1, \beta_2, \cdots, \beta_{2k}\} \in \{0,1\}^{2k}$. The maximum likelihood decoders (MLD) address this degeneracy by considering the most likely logical sector given a syndrome. In essence, an MLD can be viewed as a mapping from $\gamma$ to $\beta$. 
Neural networks can be employed to learn such mapping, i.e., the conditional probability $P(\beta|\gamma)$ from training data and generalize to unseen syndromes~\cite{bausch2024learning,gnd,qecgpt}. Since they are based on learning a variational distribution, neural network decoders do not depend on the code's topology and can be readily applied to various codes. They can also incorporate additional hardware information, such as leakage. However, the decoding time of neural network decoders is still longer than the operation of quantum circuits due to the intensive operations, the latency of classical devices, and the overhead of classical electronics. This discrepancy makes the application of learning-based decoding to real-time scenarios challenging.


\paragraph{Decoding using a quantum circuit---}
We propose to address the efficiency issue of learning-based decoding by employing a variational decoding quantum circuit. This decoding circuit incorporates parameters that we can optimize to make the probability distribution related to the final state of the circuit closely approximate $ P(\beta|\gamma) $. The learning phase is akin to training a neural network model and can be executed using the tensor network method~\cite{pan2022simulation, pan_solving_2022}. Once the learning process is complete, the quantum circuit can be deployed on a quantum device to efficiently sample a logical operator $\beta$ from the final state of the decoding circuit, given a syndrome $\gamma$.

To illustrate our scheme, consider the example depicted in Fig.~\ref{fig:circuit}. The top panel demonstrates how quantum codes with various noise models can be mapped to Tanner graphs (as shown in the top middle). The top left panel illustrates a surface code under qubit noise, where physical qubits correspond to circles and stabilizers to colored squares in the Tanner graph. The top right panel shows a distance-3 repetition code quantum memory under circuit-level noise, where error mechanisms map to circles and detectors~\cite{gidney2021Stim} to colored squares in the middle panel. The measurement outcomes in the Tanner graph generate a syndrome, which is subsequently used to decode the logical sector and apply corrections.

The decoding circuit is depicted in the bottom panel. Here, syndromes are encoded as parameters of single-qubit gates, and the operations of the quantum circuit involve parameterized single-qubit gates and non-parameterized two-qubit gates (e.g., the Controlled-Not gate), culminating in a measurement that indicates the logical sector $\beta$. In the bottom panel of Fig.~\ref{fig:circuit}, the decoding circuit comprises $ Q = 3 $ qubits and $ B = 2 $ blocks of unitary operations. Each block consists of a group of rotation $ X $ gates $\{R^X_{q,b,i}\}$ followed by a group of rotation $ Y $ gates $\{R^Y_{q,b,x_i}\}$, where $ q \in \{1,2,3\} $ denotes the qubit index in the decoding circuit, $ b \in \{1,2\} $ indicates the block index, and $ i $ is the syndrome index. The parameters of the rotation gates are determined by the product of a learnable parameter $\theta$ and the syndrome $\gamma$. The matrix form of the $ X $-rotation gates is
\[ 
R^X_{q,b,i}=
\left[ \begin{matrix} 
\cos\frac{\theta_{q,b}\cdot\gamma_i}{2}&-i\sin\frac{\theta_{q,b}\cdot\gamma_i} {2} \\-i\sin\frac{\theta_{q,b}\cdot\gamma_i} {2}&\cos\frac{\theta_{q,b}\cdot\gamma_i}{2}
\end{matrix}\right],\]
and that of the Y-rotation gates is 
\[
R^Y_{q,b,i}=
\left[ \begin{matrix} 
\cos\frac{\phi_{q,b}\cdot\gamma_i}{2}&-\sin\frac{\phi_{q,b}\cdot\gamma_i} {2} \\\sin\frac{\phi_{q,b}\cdot\gamma_i} {2}&\cos\frac{\phi_{q,b}\cdot\gamma_i}{2}
\end{matrix}.
\right]
\]
Here, $\theta$ and $\phi$ are learnable parameters; the total number of parameters is $2QBm$. In practice, one can construct a more complex circuit ans\"atze using other types of parameterized gates, more qubits, and more blocks of operations.

Once the parameters are properly learned, the circuit can perform decoding by running it on a quantum device. Specifically, after applying the single-qubit gates and two-qubit gates to the initial state, the final state is prepared and then measured in the computational basis. This process effectively yields the probability distribution corresponding to the final state in the computational basis $|\psi(\beta_1, \beta_2, \cdots, \beta_Q)|^2$, which represents the conditional probability of a bit string $\{\beta_1, \beta_2, \cdots, \beta_Q\}$ given the syndrome $\gamma$, denoted as $q(\beta|\gamma)$.

When $Q > k$, meaning the number of qubits $Q$ in the decoding circuit exceeds the number of logical qubits $k$, we can measure or select a subset of the decoding qubits. For instance, in the example of quantum memory with $k=1$ as illustrated in Fig.~\ref{fig:circuit}, we only need to measure one qubit in the decoding circuit to obtain a sample of the logical sector from $q(\beta|\gamma_1, \gamma_2, \cdots, \gamma_m)$. Optionally, we can repeat the process several times to estimate the conditional probability of the logical sector given the syndrome, then choose a logical sector that maximizes the conditional probability, rather than sampling from it.

The structure of the decoding circuit is not tied to a specific code, making it applicable to various codes and noise models. In addition to its versatility, the proposed circuit-based decoder offers several advantages:

\noindent 1. \textit{Decoding speed:} The operation time of the decoding circuit is comparable to the running time of the operation circuit. Most of the operations are single-qubit operations, whose duration is approximately $30$ ns on superconducting qubits~\cite{googlequantumai2023Suppressing}. Additionally, when the syndrome of the $i$-th measurement is $\gamma_i = 0$, the X and Y rotations are effectively identity operations, meaning we do not need to apply the unitary operation. This implies that in the decoding, only a small fraction of the single-qubit gates will be applied. This significantly reduces the operation time, especially when the syndrome is sparse (e.g., for detector error models with detectors defined using the XOR of consecutive ancilla measurements~\cite{gidney2021Stim}).

\noindent 2. \textit{Quantum sampling advantage:} With a large number of logical sectors, such as in high-rate codes with multiple qubits, generating a logical sector given a syndrome involves sampling from a joint distribution $ P(\beta|\gamma) $, where $\beta \in \{0,1\}^Q$ resides in a high-dimensional space. Classical sampling from the joint distribution typically requires modeling conditional probabilities for each variable $ P(\beta_i|\beta_{<i}, \gamma) $ using autoregressive neural networks and sampling variables sequentially~\cite{qecgpt}. However, in the proposed circuit-based decoding scheme, a configuration of the logical sector $\beta$ can be directly sampled using a quantum device, leveraging Born's rule in quantum mechanics, rather than computing conditional probabilities as in classical sampling. We have witnessed the quantum advantage for the random circuit sampling problem~\cite{arute2019quantum,zuchongzhi3}, and we anticipate a quantum speedup in sampling logical sectors for high-rate decoding problems.

\paragraph{Training ---}
The training of the decoding circuit involves iteratively adjusting the parameters of the single-qubit gates to make the conditional distribution $ q(\beta|\gamma) $ as close as possible to the distribution observed in the training data. The data comprises pairs of $\gamma$ and $\beta$, which can be derived from an error model or gathered experimentally. Similar to neural network decoders, the training process for the decoding circuit includes both forward and backward passes. During the forward pass, we construct the single-qubit gates using the training syndrome $\gamma$ and the parameters $\theta$ and $\phi$. We then simulate the quantum circuit using tensor networks to obtain the final state and compute the conditional probability $ p(\beta|\gamma) $ as the output. Subsequently, we formulate the cross-entropy loss function based on this output and the corresponding label $\gamma$ from the training data. In the backward pass, we utilize the backpropagation algorithm to compute the gradients of the loss function with respect to the parameters. These gradients are then employed by an optimizer, such as Adam, to update the parameters.
\begin{figure*}[htb]
    \centering
    \includegraphics[width=0.32\linewidth]{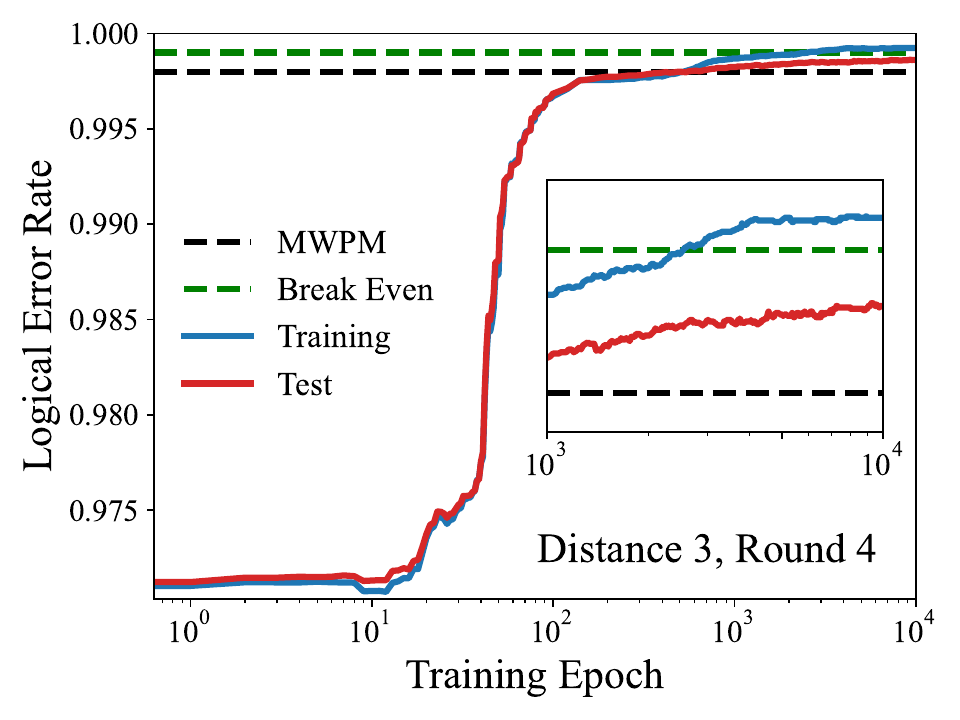}
    \includegraphics[width=0.32\linewidth]{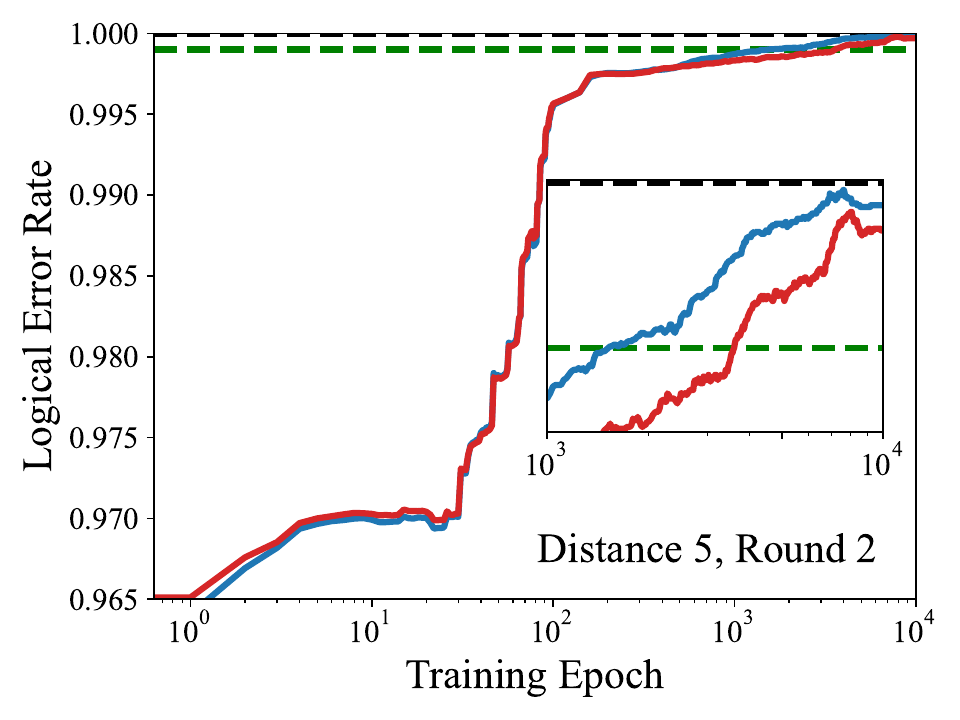}
    \includegraphics[width=0.32\linewidth]{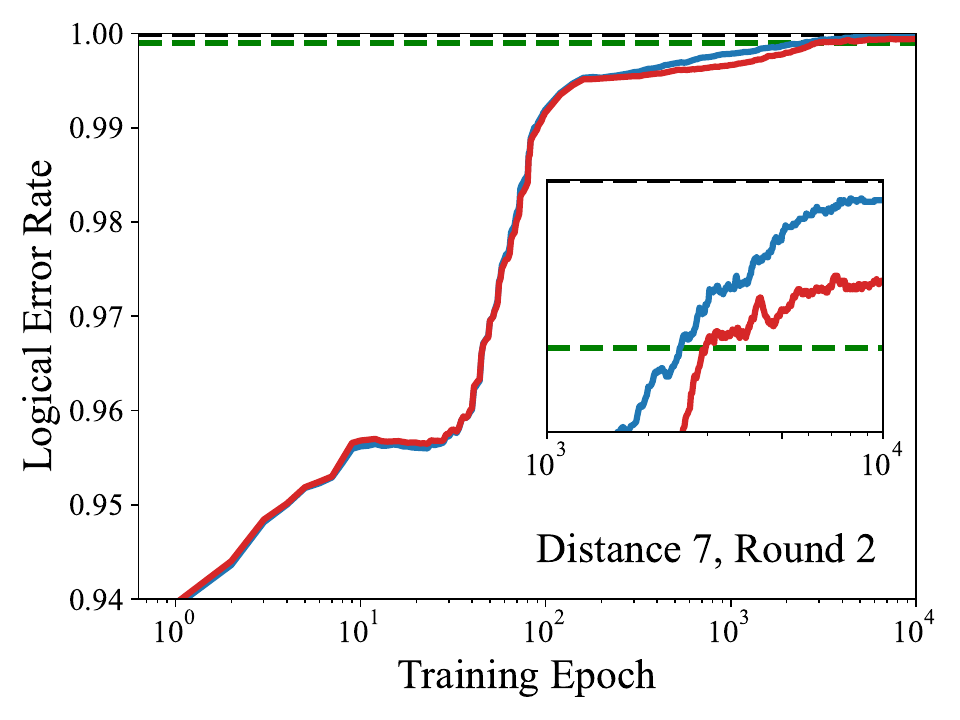}
    \caption{
   Numerical results for our quantum circuit decoder on the surface code quantum memory in the Z-basis under depolarizing circuit-level noise with various distances and rounds. The physical error rates are fixed at $0.001$. The evolution of the logical error rates on both training and test syndromes is compared with the minimum weight perfect matching results on the test syndromes and the break-even point with a $0.001$ logical error rate. The training dataset comprises $200,000$ syndromes, while the test dataset contains $100,000$ syndromes for each panel. The decoding circuits are composed of $3$ qubits and $10$ blocks.}
    \label{fig:sur}
\end{figure*}
\paragraph{Numerical experiments---}
To validate whether the proposed circuit decoder can effectively suppress errors in practical quantum codes, we conducted numerical experiments using the surface code quantum memory in the Z-basis under depolarizing circuit-level noise. We tested the surface code with distances 3, 5, and 7~\cite{acharya_quantum_2025}. The evolution of the logical error rate given by our decoding circuit with training steps is shown in Fig.~\ref{fig:sur} and compared with the MWPM algorithm. In the experiments, we fixed the physical error rate to $0.001$ to mimic the error level of current superconducting quantum devices. The decoding circuits for each code are composed of 3 qubits and 10 blocks of operations, and we consider the decoding circuit to be noiseless. We trained all the decoding circuits using 200,000 syndromes randomly generated from the error model using the Stim library~\cite{gidney2021Stim}. We also reported 100,000 test syndromes randomly generated using Stim with a different random seed.

From the figure, we observe that as the number of training epochs increases, the logical error rate for both training syndromes and test syndromes initially decreases from a high value to a low value. For the surface code with distances 3 and 4 rounds of measurements, the test logical error rate after 10,000 training epochs is better than that of MWPM, while still lower than the break-even point $0.999$. This is attributed to the short distance of the code. For distances 5 and 7, we can clearly see that after $10^4$ training epochs, the test logical error rate exceeds the break-even value.

\paragraph{Self-correcting quantum computer---} 
We have introduced a method for performing error correction using measured classical syndromes from the operation quantum circuit, facilitated by a decoding quantum circuit. In this framework, the decoding circuit is connected to the operation circuit through measurements and classical electronics. The measurement outcomes are utilized to parameterize the learned variational single-qubit gates, which execute the decoding process. This approach redefines the decoding task as a quantum task rather than a classical one.

A direct implication of this redefinition is that measuring ancilla qubits to obtain classical syndromes for error detection and correction is no longer necessary. Instead, the state of the ancilla qubit (e.g., whether it flipped from $|0\rangle$ to $|1\rangle$ in a Z-basis measurement) can directly control the parameters of the decoding circuit (e.g., through a controlled single-qubit gate with learned parameters) to perform decoding. Additionally, the outcome of the decoding circuit can be directly fed back to the operation circuit using controlled gates to perform logical operations. We refer to this scheme as the \textit{self-correcting circuit}.

\begin{figure}[htb]
     \centering
\includegraphics[width=0.8\linewidth]{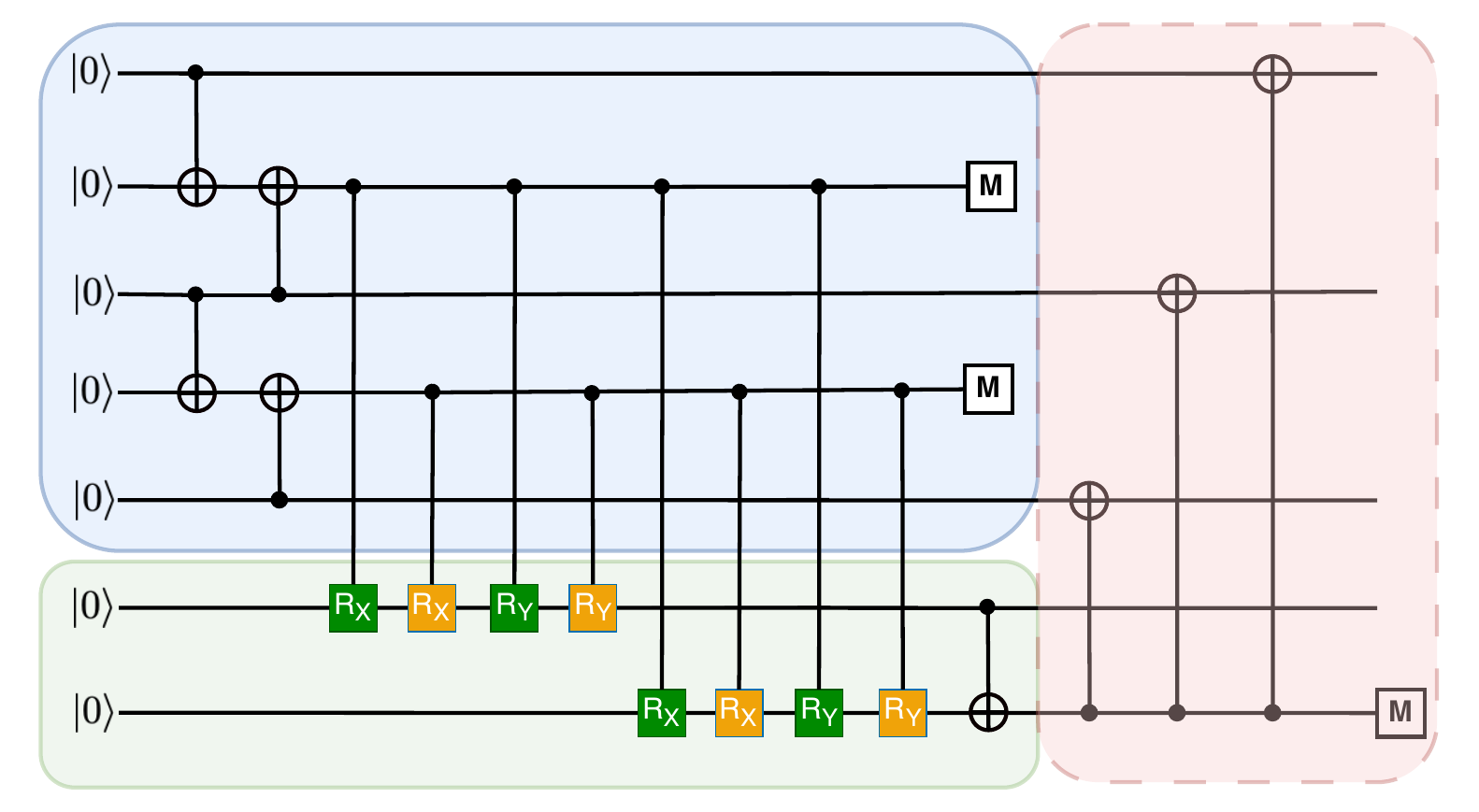}
     \caption{Illustration of the self-correcting repetition-code quantum memory with distance $3$. The blue shaded part of the circuit represents the memory section, comprising three data qubits and two ancilla qubits. In this round, the ancilla qubits are measured. These ancilla qubits are directly utilized to govern the learned single-qubit gates in the decoding segment of the circuit (depicted in green), which encompasses $2$ decoding qubits. The second decoding qubit is ultimately employed to manipulate the data qubits, introducing a controlled logical X gate (depicted in red) to rectify the error within the repetition memory.}
     \label{fig:self}
 \end{figure}

In Fig.~\ref{fig:self}, we provide an illustrative example. The circuit includes a repetition code with 3 data qubits, 2 ancilla qubits, and 2 decoding qubits. The single-qubit gates in the decoding part (green shaded area) are trained using a classical algorithm and are directly controlled by the two ancilla qubits. After the operation of the learned single-qubit gates, the second decoding qubit conditionally adds a logical X operation (the red shaded area) to the three data qubits. If the decoding part were noiseless, the results would closely mirror the performance of the decoding circuit we demonstrated in the preceding sections. However, it is important to note that this example differs from the decoding circuit because it does not involve any classical operations. The measurement outcomes are not utilized as classical inputs to the decoder, and the decoder's output is not classical. Rather, the state is used to determine whether to apply a logical X operation to the circuit.
\paragraph{Discussions--}
We have introduced a method that employs a variational quantum circuit to decode a quantum error-correcting code. The proposed method is based on learning and is applicable to various codes under different kinds of noise. The decoding time of the proposed method is aligned with the operation time of the noisy quantum circuit. This enables the method to correct logical errors in the noisy quantum circuit on the fly.  

As a proof-of-concept, we have demonstrated our results using numerical experiments with only 3 decoding qubits, which already provided accuracy comparable to the minimum-weight perfect matching algorithm on the surface codes under circuit-level noise. Using more decoding qubits, more blocks of operations, and a better circuit ans\"atze can significantly enhance the representational power of the circuit and achieve even smaller logical error rates. So far, we have performed learning using tensor networks, but in principle, one can use a quantum device to perform sampling and employ classical optimization methods for learning. 

In the experiments, we have only considered the noiseless decoding circuit, which is assumed to be fault-tolerant. A noisy decoding circuit could also accomplish the decoding task, as the decoding process itself is probabilistic. Moreover, the decoding circuit is much less noisy than the circuit to be corrected because it has fewer qubits and is primarily composed of single-qubit gates without reset and measurements. In practical noisy decoding devices, one could also consider the noisy decoding circuit as an overall ans\"atze for decoding and learning using the real quantum device. We will explore this in future work.

Our method demonstrates that decoding of quantum error-correcting codes, a task traditionally considered classical, can be achieved using a quantum circuit. This suggests that a noisy quantum computer may correct itself using part of its own hardware, without relying on classical measurements and processing. In this sense, the fault-tolerant quantum computer may be capable of self-correction.
\begin{acknowledgements}
This work is supported by Projects 12325501 and 12447101 of the National Natural Science Foundation of China. P.Z. acknowledges Hanyan Cao, Dongyang Feng, Feng Pan, and Keyang Chen for helpful discussions at the early stage of this work. A Python implementation of the proposed algorithm can be found at https://github.com/panzhang83/dqc.
\end{acknowledgements}
%

\end{document}